\documentclass{aip-cp}
\usepackage[numbers,elide]{natbib}
\usepackage[T1]{fontenc}
\usepackage{amssymb,amsmath,parskip}

\usepackage{graphicx,txfonts}

\begin{document}
\title{Correlations within the Non-Equilibrium Green's Function Method}

\author[aff1]{M.H. Mahzoon}
\eaddress{hossein@pa.msu.edu}
\author[aff1,aff2]{P. Danielewicz}
\author[aff3]{A. Rios}

\affil[aff1]{Department of Physics and Astronomy, Michigan State University, East Lansing, MI 48824-1321, USA}
\affil[aff2]{National Superconducting Cyclotron Laboratory, Michigan State University, East Lansing, MI 48824-1321, USA}
\affil[aff3]{Department of Physics, University of Surrey, Guildford, Surrey GU2 7XH, United Kingdom}

\maketitle

\begin{abstract}
Non-equilibrium Green's Function (NGF) method is a powerful tool for studying the evolution of quantum many-body systems. Different types of correlations can be systematically incorporated within the formalism. The time evolution of the single-particle Green's functions is described in terms of the Kadanoff-Baym equations. The current work initially focuses on introducing the correlations within infinite nuclear matter in one dimension and then in a finite system in the NGF approach. Starting from the harmonic oscillator Hamiltonian, by switching on adiabatically the mean-field and correlations simultaneously, a correlated state with ground-state characteristics is arrived at within the NGF method.  Furthermore the use of cooling to for improving the adiabatic switching is explored.
\end{abstract}

\section{Introduction}
The ultimate goal of this project is to investigate nuclear collisions following the Non-equilibrium Green's Function (NGF) method within space-time representation. To initialize a many-body system, the correlated ground-state needs to be constructed, static under real-time evolution.  The latter can be ensured by employing the same infrastructure for correlating the system as for the evolution, and by adiabatically switching on the correlations together with the mean field.  We originally trap the finite system in a harmonic oscillator potential and then adiabatically switch off that potential while ramping up the mean-field in combination with a Gaussian residual interaction.  The conversion of the interactions can be aided with a transient cooling friction term.

\section{Kadanoff-Baym Equations}
One-body information within a many-body system is accounted for in terms of the one-body density matrix.  In the mean-field approximation, the dynamics of a many-body system can be cast in a self-consistent and self-contained manner in terms of that one-body density matrix.  Provided that the effects of short-range correlations are short-lived, the dynamics of a many-body system can be approximately described in a self-contained manner in terms of the single-particle Green's functions that are two-time extensions of the density matrix, cf.~\cite{Danielewicz1984}:
\begin{eqnarray}
\label{g_less}
G^<(x_1 \, t_1;x_{1'} \, t_{1'}) & = &  i \langle \hat \phi^{\dagger}(x_{1'} \, t_{1'}) \; \hat \phi(x_1 \, t_1)\rangle \, , \\
G^>(x_1 \, t_1;x_{1'} \, t_{1'}) &  = & -i \langle \hat \phi (x_1 \, t_1) \; \hat \phi^\dagger(x_{1'} \, t_{1'})\rangle \, .
\end{eqnarray}
Here, $\hat \phi (x \, t)$ annihilates a particle at location $x$ at time $t$ and the expectation value $\langle \cdot \rangle$ is with respect to an uncorrelated state at some starting time $t_0$.  With space and time arguments lumped together as $(x_1 \, t_1) \equiv 1$, to simplify notation, the Green's functions can be shown to satisfy the Kadanoff-Baym (KB) equations~\cite{kadanoff}:
\begin{equation}
\begin{aligned}
\left [ i\hbar \frac{\partial}{\partial t_1} + \frac{\hbar^2}{2m} \frac{\partial^2}{\partial x_1^2} \right ] G^{\gtrless} (1 \, 1')= \int \text{d}x_{\bar 1} \, \Sigma_{HF}(1\bar 1) \, G^{\gtrless} (\bar 1 1')+ & \int_{t_0}^{t_1} \text{d}{\bar 1} \left [\Sigma^>(1 \, \bar 1 )-\Sigma^<(1 \, \bar 1 ) \right ]G^{\gtrless} (\bar 1 \, 1') \\
- & \int_{t_0}^{t_{1'}} \text{d}{\bar 1} \, \Sigma^{\gtrless}(1 \, \bar 1 ) \left [ G^> (\bar 1 \, 1')- G^< (\bar 1 \, 1') \right ] \, ,
\end{aligned}
\end{equation}
\begin{equation}
\begin{aligned}
\left [ -i\hbar \frac{\partial}{\partial t_1'} + \frac{\hbar^2}{2m} \frac{\partial^2}{\partial x_1'^2} \right ] G^{\gtrless} (1 \, 1')= \int \text{d}x_{\bar 1}  \, G^{\gtrless} ( 1 \, \bar 1 ) \, \Sigma_{HF}(\bar 1 \, 1') \;+ &
\int_{t_0}^{t_1} \text{d}{\bar 1} \, \left [G^>(1 \, \bar 1 )-G^<(1 \, \bar 1 ) \right ]   \Sigma^{\gtrless} (\bar 1  \, 1')  \\
- &\int_{t_0}^{t_{1'}} \text{d}{\bar 1} \, G^{\gtrless}(1 \, \bar 1 ) \left [ \Sigma^> (\bar 1 \, 1')- \Sigma^< (\bar 1 \, 1') \right ] \, .
\end{aligned}
\end{equation}
Here, $\Sigma_{HF}$ and $\Sigma^{\gtrless}$ are the self-energies that can be approximated in terms of the single-particle Green's functions $G^{\gtrless}$.  The self-energy $\Sigma_{HF}$ accounts for the most direct mean-field effects and the energies $\Sigma^{\gtrless}$ incorporate the effects of correlations.  When the self-energies are consistently approximated, the KB equations preserve the conservation laws of particle number, momentum and energy.

In the past, we explored the mean-field dynamics in the context of employing the KB equations for describing nuclear collisions~\cite{Rios2011}.  We examined arriving at the mean-field ground-state through adiabatic switching and collisions of nuclear slabs.  Here, we investigate the process of switching on the correlations for infinite matter and for slabs.

\section{Importance of Initial-State Preparation}
\label{adiabatic}

In the previous sections the KB equations which govern the time evolution of a many-body system were introduced. Solving the latter equations yields one-body information about the system and also many-body provided the effects of correlations are short-lived.  In studying collisions, nuclear systems need to be initialized consistently with the KB equations used to follow the collisions.  Thus, if an uncorrelated initial state is naively followed with KB equations incorporating correlations, then that system may spew particles and even explode and/or violently oscillate ahead of any collision.  In the previous work~\cite{Rios2011}, we employed adiabatic switching to arrive at mean-field initial states consistent with a mean-field version of the KB equations.  Examples where adiabatic switching was used for constructing correlated many-body states include Refs.~\cite{tohyama_stationary_1994,pfitzner_vibrations_1994}.  In \cite{Danielewicz1984}, imaginary-time evolution was employed, requiring development of a separate computational infrastructure, but equivalent to the adiabatic switching under proper circumstances.

\section{Employed Interactions}

\begin{figure}[h]
  \centerline{
  \includegraphics[width=.55\textwidth]{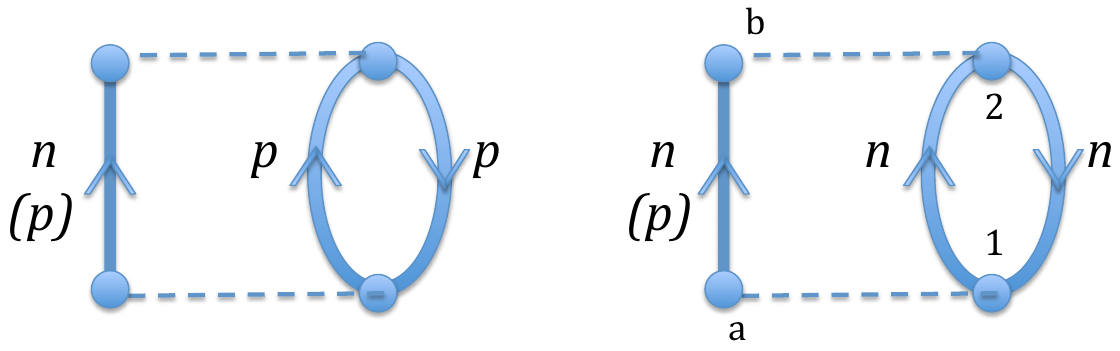}}
  \caption{Direct Born diagram contributions to the self energy for neutrons (protons).}
  \label{fig:borniso}
\end{figure}

In the Hartree-Fock part of the self-energy, we use a local Skyrme-type interaction such as in Ref.~\cite{Rios2011}.  For the self-energies $\Sigma{\gtrless}$, we employ the so-called self-consistent Born diagram illustrated in Fig.~\ref{fig:borniso}, yielding
\begin{equation}
\Sigma^{\gtrless} (p \, t;p' \, t') =  \int \frac{\text{d} p_1}{2\pi}\frac{\text{d} p_2}{2\pi} \; V(p-p_1) \, V(p'-p_2)\; G^{\gtrless}(p_1 \, t;p_2 \, t')\, \Pi^{\gtrless}(p-p_1 \, t; p'-p_2 \, t') \, ,
\end{equation}
where
\begin{equation}
\Pi^{\gtrless}(p \, t;p' \, t') =  \int \frac{\text{d}p_1}{2\pi}\frac{\text{d}p_2}{2\pi}G^{\gtrless}(p_1 \, t;p_2 \, t') \, G^{\gtrless}(p_2-p' \, t';p_1-p \, t) \, .
\end{equation}

In the semiclassical low-density limit, the self-energies $\Sigma{\gtrless}$ represent phase-space feeding and depletion rates where cross sections are described in the Born approximation in terms of the employed residual interaction.  To describe semi-quantitatively 3D rates within the 1D calculation, we employ a residual interaction modulated by a Gaussian:
\begin{equation}
\label{eq:Vres}
\begin{aligned}
V(p) =& V_0 \, \sqrt{\pi} \, \eta^2 \, p^2 \, \text{e}^{-{\eta^2 \, p^2}/{4}} \, , \\
V(x) =& V_0 \, \left ( 1-2\frac{x^2}{\eta^2} \right ) \, \text{e}^{-{x^2}/{\eta^2}} \, .
\end{aligned}
\end{equation}
The constants $V_0$ and $\eta$ are adjusted to yield reasonable cross sections and quasiparticle content of single-particle states in the ground state.

\section{Adiabatic Transformation of Interactions}

In the case of a finite system we generally start with one that is uncorrelated and trapped in a harmonic oscillator potential $U_0(x) = \frac 1 2 m \Omega^2 x^2$.  One way to choose the optimal  frequency $\omega$ for the potential is by demanding that the r.m.s.\ radius for the trapped matter is the same as for a self-bound uniform slab at normal density~\cite{Rios2011,ring}.  In the case when only the self-consistent mean field $\Sigma_\text{HF}$ is active for the self-bound slab, the adiabatically converted potential may be represented as
\begin{equation}
U_t (x \, t) = F(t)\, U_0(x) +[1-F(t)] \, \Sigma_\text{HF} (x \, t) \, .
\end{equation}
Here $F(t)$ is a switching function that changes between 1 and 0 around some switching time $t_s$.

In the literature there had been discussions of optimal choice of switching functions $f$ for adiabatic switching~\cite{Watanabe}.  If the switching interval is from $t_0$ to $t_1$, we ensure that $F$ changes between 1 and 0 there by taking
\begin{equation}
F(t)=\frac{f(t-t_s)-f(t_1-t_s)}{f(t_0-t_s)-f(t_1-t_s)} \, ,
\end{equation}
and our standard function $f$ is~\cite{Rios2011}
\begin{equation}\label{eq:f}
f(t)=\frac{1}{1+e^{t/\tau}}
\end{equation}
where $\tau$ controls the pace of adiabatic switching.

\begin{figure}[h]
  \centerline{\includegraphics[width=.7\textwidth]{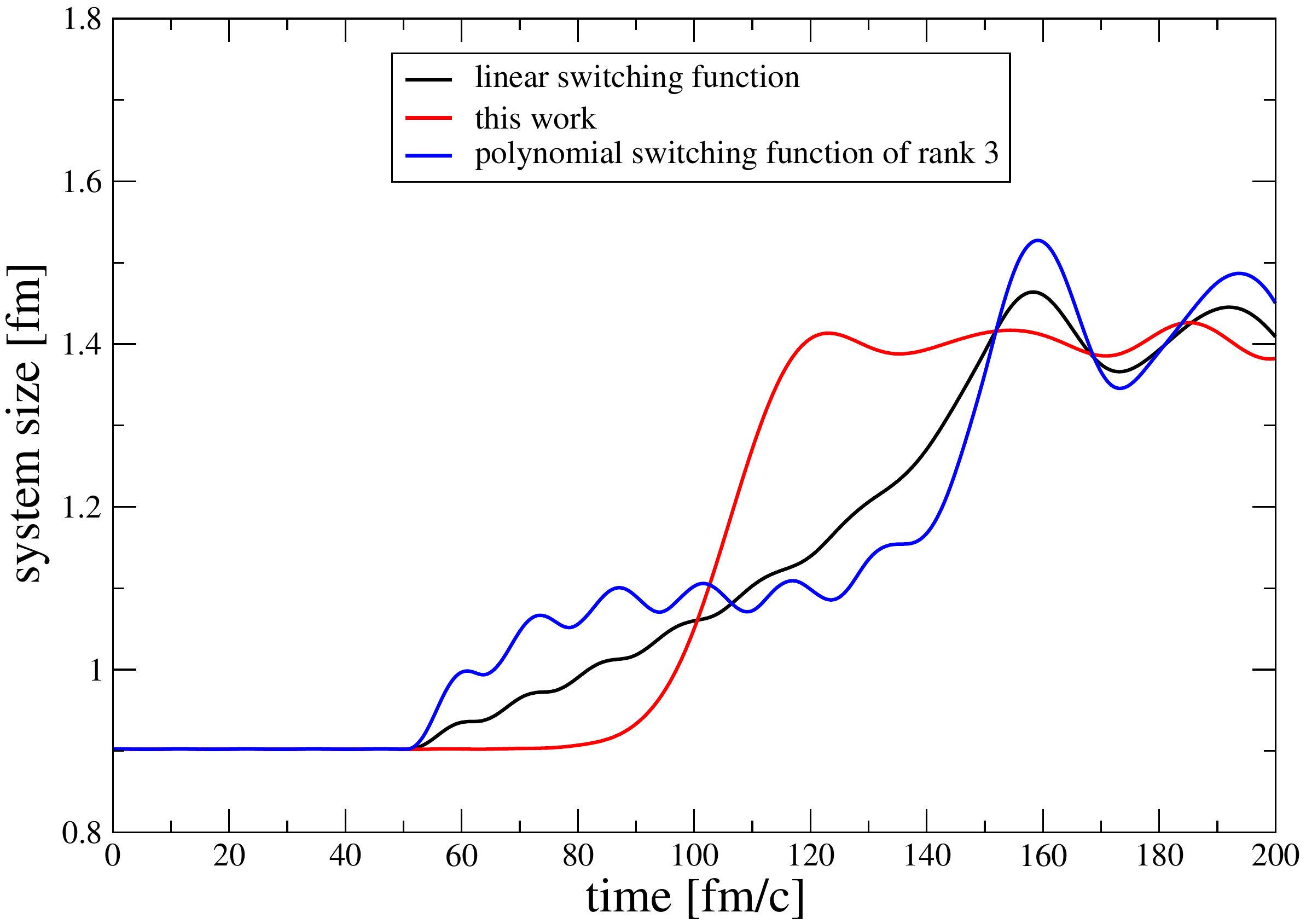}}
  \caption{Time evolution of the size of an $A=4$ system for different types of switching functions, with the switching active in the time interval of $[50,150] \, \text{fm}/c$. }
  \label{fig:switchings}
\end{figure}

Given sufficiently long switching time, desirable properties of an interacting initial state for a collision can be reached, such as low net energy, sharply ending density distribution confirming low temperature, and little time dependence.  Due to the 2-time structure of NGF, though, a prolonged switching on of the interactions can be quite costly computationally.  With this, it is of advantage to arrive at the desirable properties over as short time as possible, optimizing the adiabatic path.  In Ref.~\cite{Watanabe} different switching functions $f$ were proposed that we test in Fig.~\ref{fig:switchings} in the case of switching of the pure mean field of Ref.~\cite{Rios2011} for a nuclear slab.  We show there system size in terms of $\langle |x| \rangle$ in the case of the same $U_0$ and the same final mean-field.  Short switching times lead to oscillating mean-field states with the amplitude of oscillations diminishing as the pace of switching slows down.  It is apparent that our original function ("this work"), Eq.~\eqref{eq:f}, performs well against other proposed switching functions~\cite{Watanabe}, allowing to arrive at the desirable properties of the interacting initial state in a shorter time than for the other proposed functions.

\section{Infinite Nuclear Matter}

While working in 1D, we want to tie as closely as possible the properties of 1D systems to those of slabs in 3D, uniform in two directions.  Through the use of Hugenholtz-van Hove theorem, we have shown how to transcribe density and energy between the systems.  For the density the relation is \begin{equation}
n_{3D}=\xi \;  n_{1D} \, , \qquad \xi = \sqrt{\frac53}\left (   \frac{\pi n_0^2}{6 \nu^2} \right)^{1/3} \, ,
\end{equation}
where $n_{1D}(x,t)= -i \nu G^< (x \, t;x \,t)$, $\nu=4$ is spin-isospin degeneracy and $n_0 = 0.16 \, \text{fm}^{-3}$ is the normal density.

In deciding on interaction parameters and exploring system properties, we first examine infinite isospin-symmetric nuclear matter.  We switch on the residual interactions adiabatically, just as in the case of mean field~\cite{Rios2011}.  The residual interactions in $\Sigma^{\gtrless}$ contribute to net energy, besides the mean-field interactions, in terms of the correlation energy.  Details, besides on interaction parameters depend on spatial grid employed in calculations, that introduces a high-momentum cut-off.  In production calculations we employ a mesh with the separation of $\text{d}x = 0.67 \, \text{fm}$.  The mesh implies what time steps can be used in the time integration, with a finer mesh requiring shorter time steps.

\begin{figure}[h]
  \centerline{\includegraphics[width=.65\textwidth]{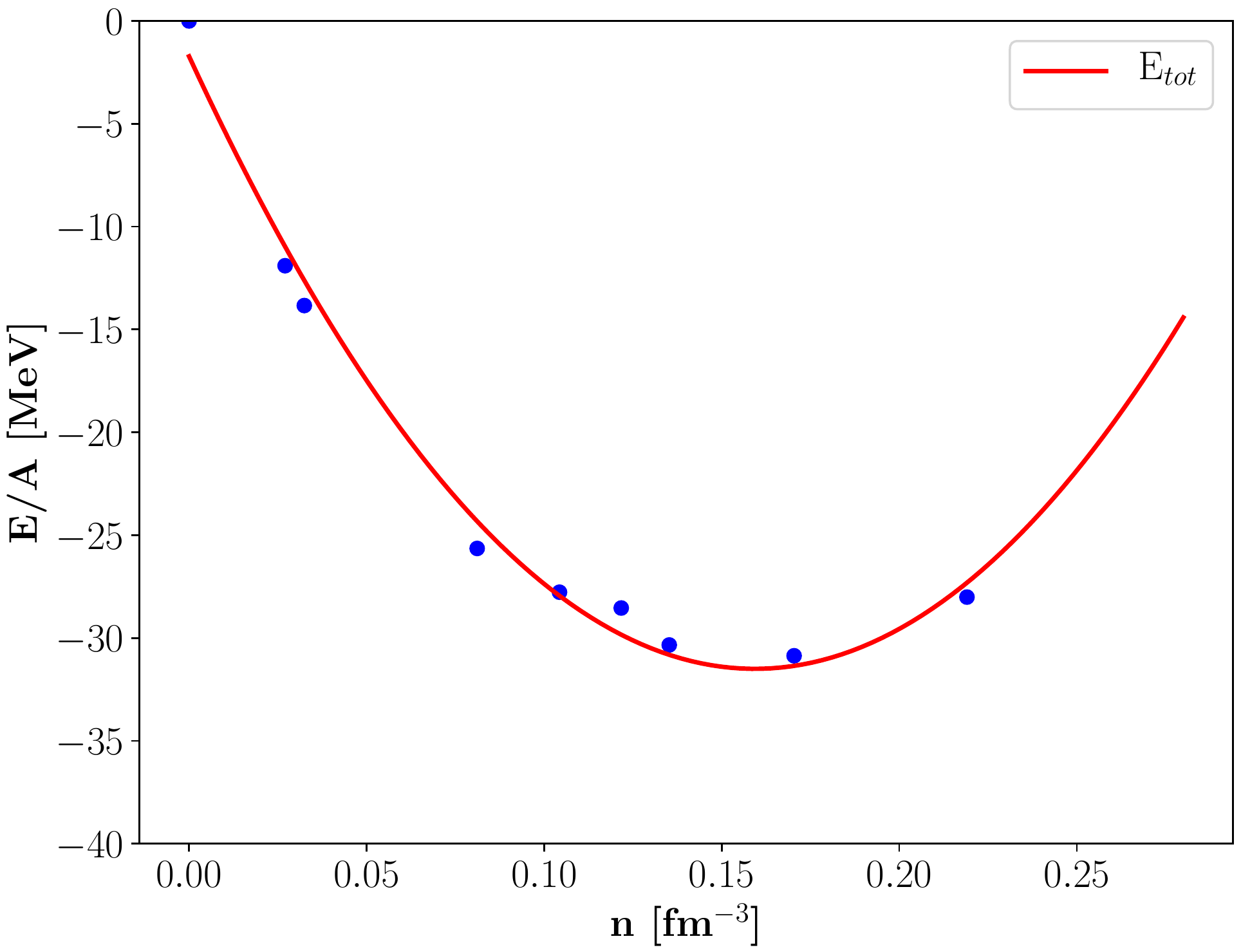}}
  \caption{Net energy per nucleon as a function of 3D density in cold correlated nuclear matter.  Symbols represent values arrived at sample densities and the line represents a smooth interpolation of the results.}
  \label{fig:e_tot_infinite}
\end{figure}

We supplement the residual interactions with mean field so that the net energy, consisting of kinetic, correlation and mean-field energy, minimizes at $n_0$, with a sensible curvature around the minimum.  That energy, in terms of points for sample densities, and in terms of a line for a smooth interpolation of the results, is shown in Fig.~\ref{fig:e_tot_infinite}.  The numerically relatively low energy values are tied to the fact that the energy in 1D excludes the kinetic energy associated with transverse degrees of freedom, there in the 3D matter, but frozen here.

\section{Finite Systems}

\begin{figure}[h]
     \centering
      \includegraphics[width=.75\textwidth]{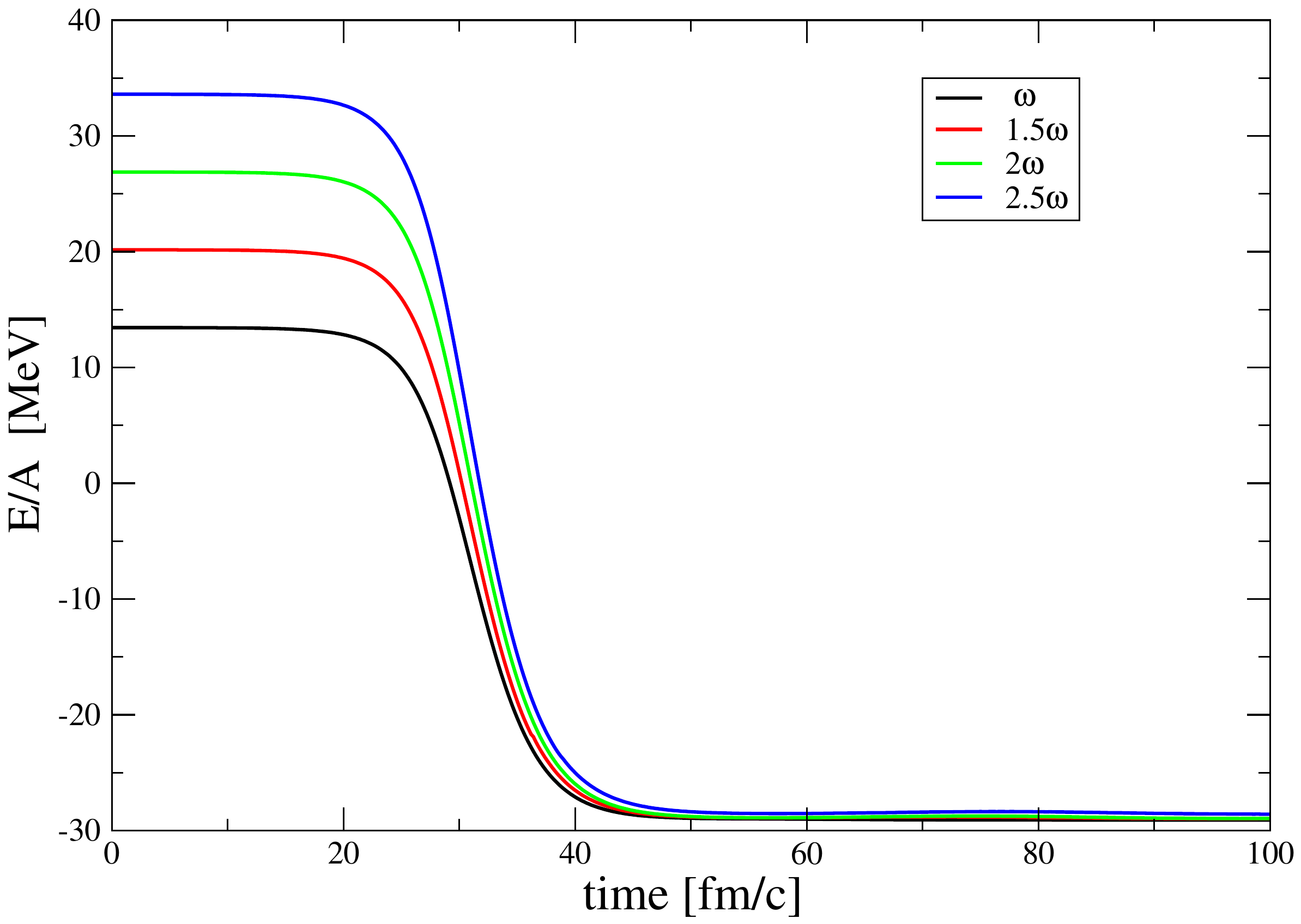}
  \caption{Net energy per nucleon as a function of time for the $A=4$ system initialized in the ground state  particle time evolution for different initial HO potential frequencies $\Omega$.}
  \label{fig:e_tot_infinite_w}
\end{figure}

Upon deciding on the combination of the residual interaction and the mean field, we proceed to studying the finite systems that are started out in a harmonic trap.  For efficient switching, the desired initial state for correlated evolution should be arrived at, no matter what the details are of the starting Hamiltonian, at least within some range of those details.  Figure~\ref{fig:e_tot_infinite_w} shows the evolution of net energy during adiabatic switching when starting an $A=8$ system in the ground state of HO potential (filled orbitals $N=0$ and 1) at different frequencies $\Omega$ in the units of the nominally optimal frequency $\omega$.   It is apparent that the systems arrive at about the same energy no matter how the systems are started.


 \begin{figure}[h]
   \centering
  \includegraphics[width=.67\textwidth]{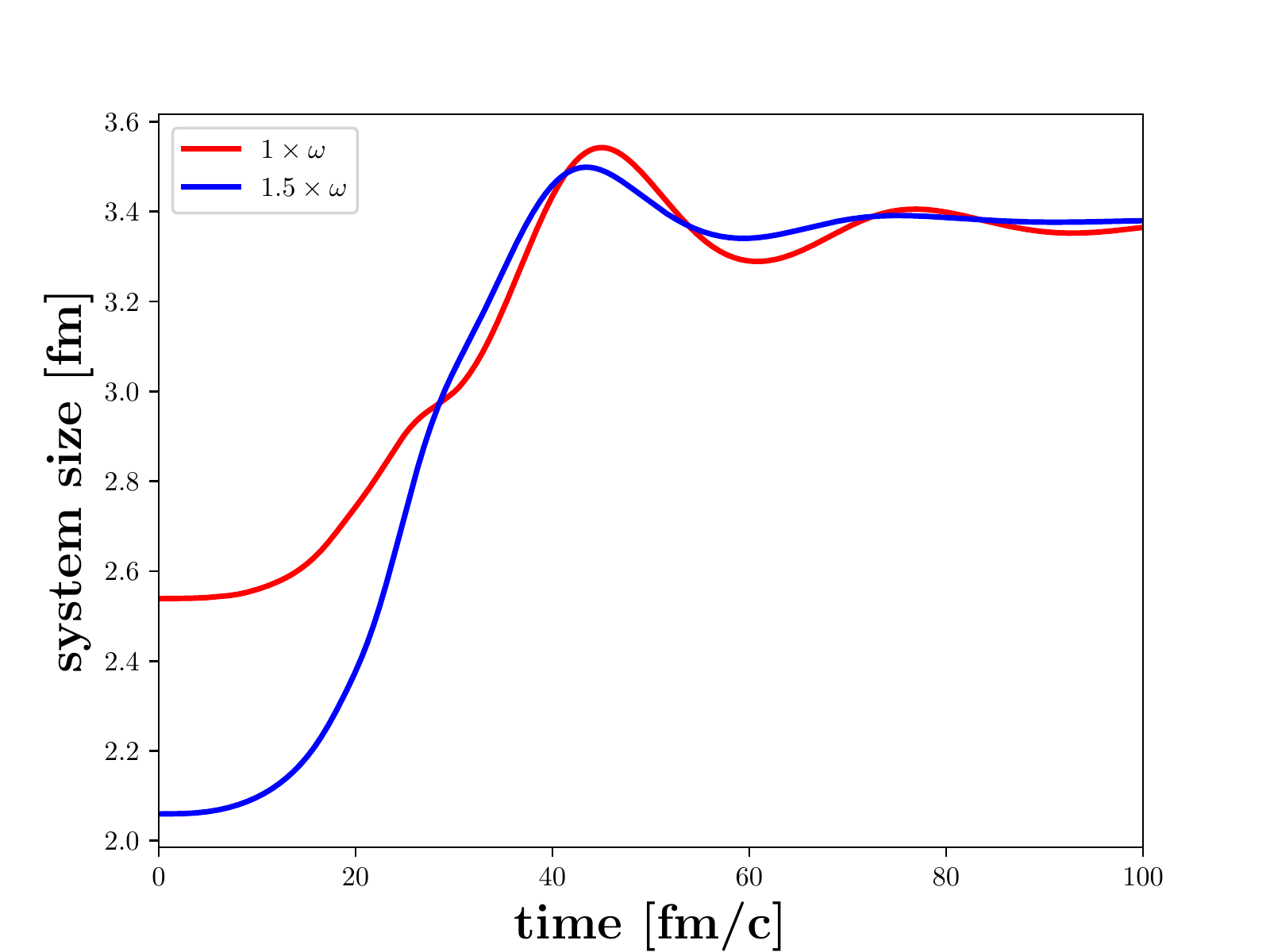}
  \caption{System size (in fm) as a function of time in the course of adiabatic transformation of interactions for different initial HO potential frequencies $\Omega$ in the units of optimal frequency $\omega$ from analytic considerations.}
  \label{fig:size_infinite_w}
\end{figure}

Near the minimum the energy per nucleon is a relatively forgiving observable and the geometric characteristics can become more telling on the arrival to the ground state irrespectively of the starting conditions.  In Fig.~\ref{fig:size_infinite_w} we show the evolution of $A=12$ ($N=0,1,2$) system size in terms of $\langle|x|\rangle$, in the course of the adiabatic transformation of interactions, for a couple of starting frequencies.  It is seen that though $\Omega = 1.5 \, \omega$ yields a worse match between the initial and final sizes, than $\Omega = \omega$, it allows to reach the desired final state a tad earlier with a more static final state.

\section{Cooling Friction}

\begin{figure}[h]
  \centerline{\includegraphics[width=.67\textwidth]{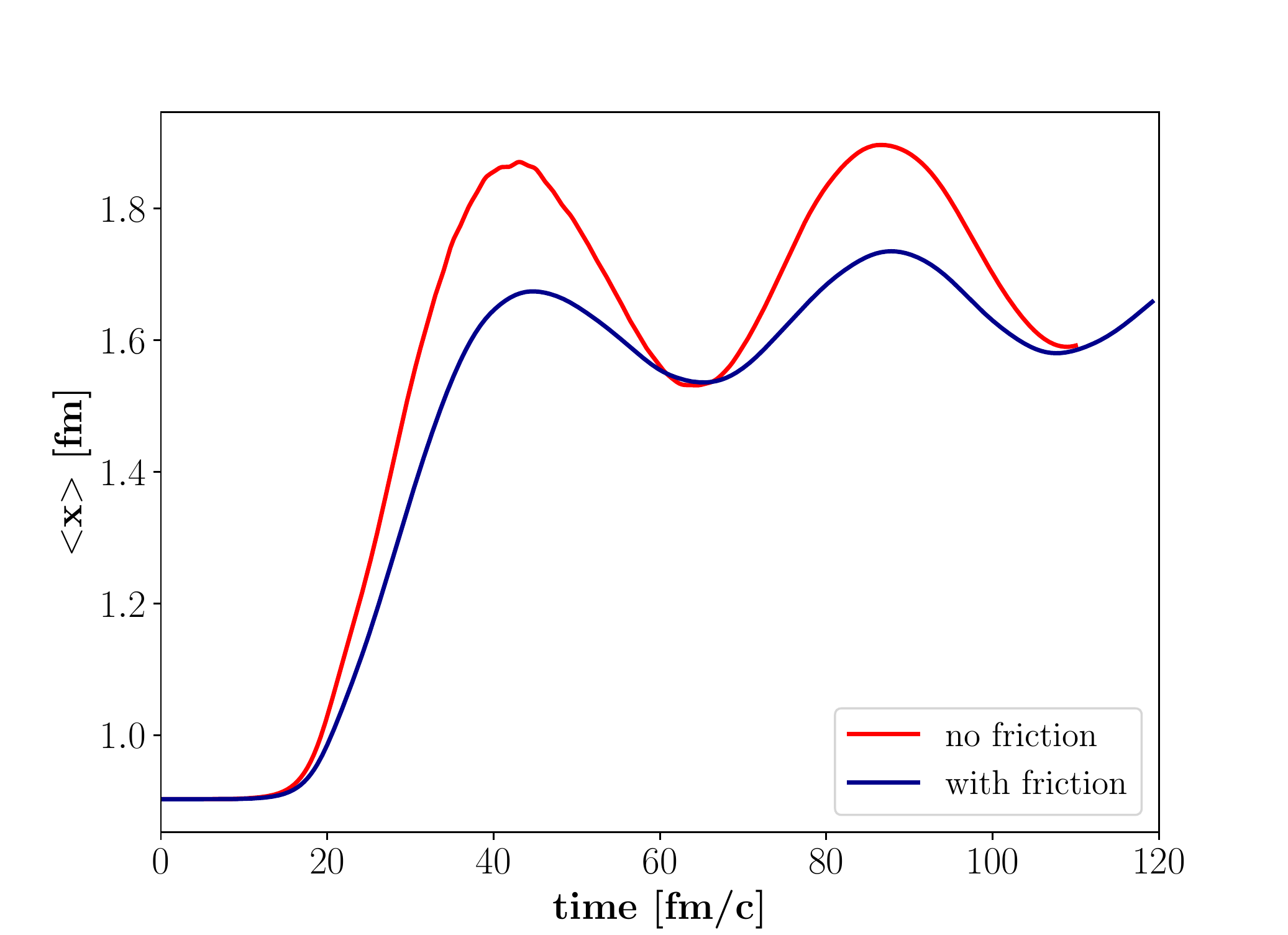}}
  \caption{Effect of the supplemental cooling-friction term on the evolution of system size when transforming the interactions.}
  \label{fig:fricyesno}
\end{figure}

Obviously, there may be more ways of transforming interactions between the initial and final states, leading to the ground-state characteristics for the latter, than through a linear combination of the initial and final interactions~\cite{bulgac,patra_shortcuts_2017}.  In particular, in~\cite{bulgac} it was proposed to supplement the potential in the equations of motion by a term breaking the time-reversibility and proportional to the time derivative of density in any particular representation.  Figure~\ref{fig:fricyesno} illustrates the impact of a friction term proportional to time derivative of density in space representation, when gradually switching on the interactions for an $A=4$ ($N=0$) system.  Outside of the switching interval, the friction is turned off.  It is apparent that the friction allows to arrive at a state that is more compact and more static than without the friction.

\begin{figure}[h]
  \centerline{\includegraphics[width=\textwidth]{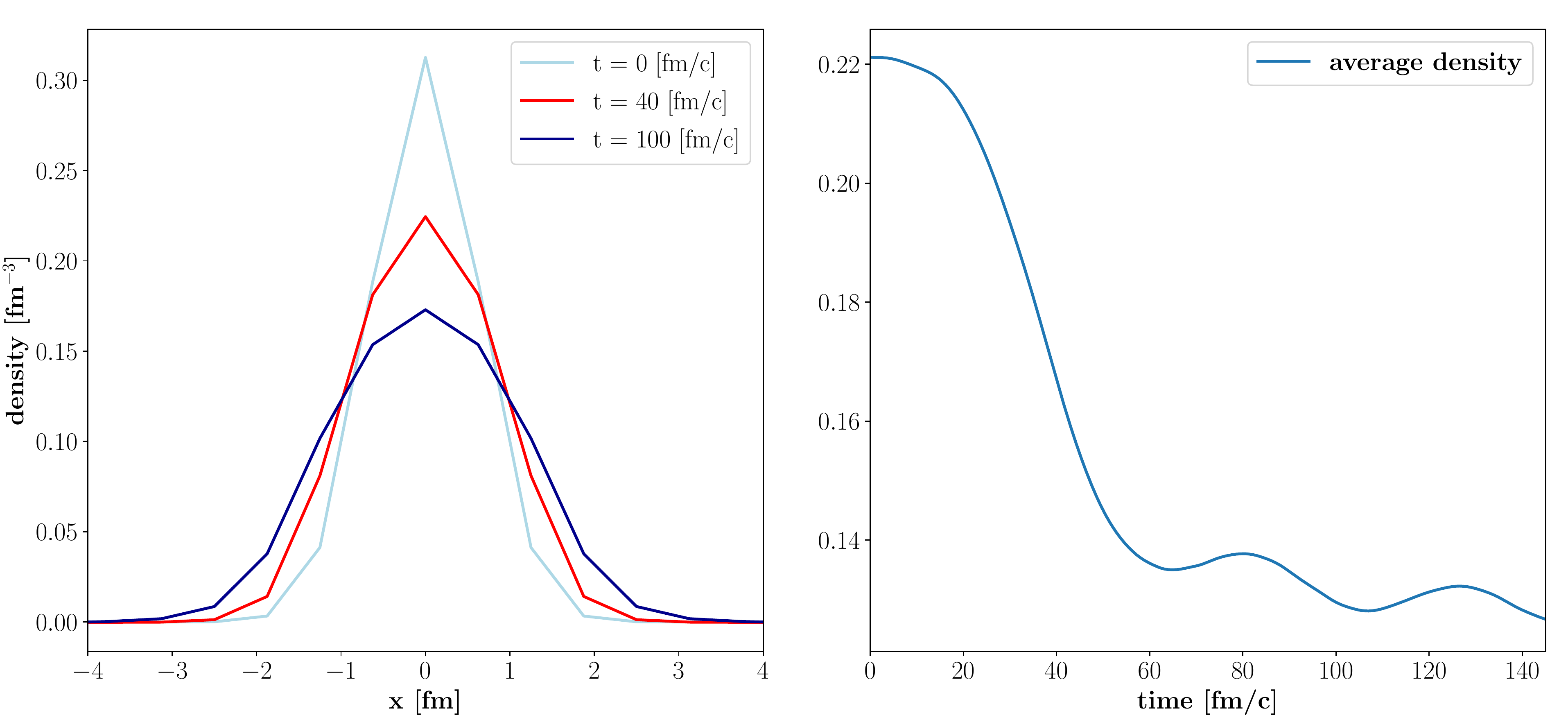}}
  \caption{Time evolution of the density in position space (left) and of the average density (right) for an $A=4$ system during and after the transformation of interactions.}
  \label{fig:av_density}
\end{figure}

\begin{figure}[h]
  \centerline{\includegraphics[width=.67\textwidth]{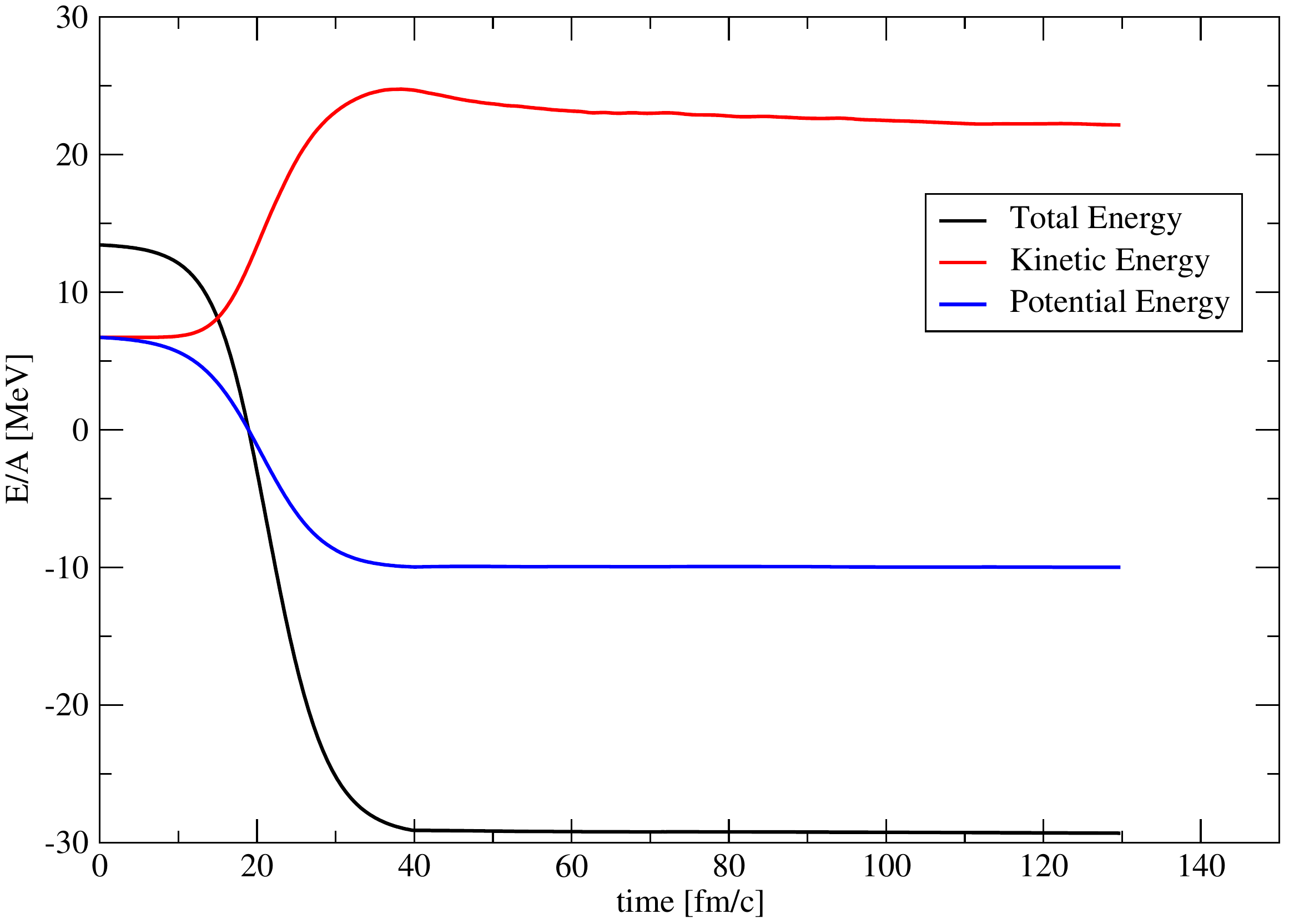}}
  \caption{Evolution of net energy per nucleon for an $A=4$ system, and its potential and kinetic components, during and after the adiabatic transformation of interactions.}
  \label{fig:etotal}
\end{figure}

The evolution of size in Fig.~\ref{fig:fricyesno} with friction is complemented with Fig.~\ref{fig:av_density} that shows the changes in the density profile during adiabatic transformation of the interactions, on the left, and the evolution of the average density, on the right.  Finally, in Fig.~\ref{fig:etotal} the evolution of net energy per nucleon is shown, for the $A=4$ slab, together with the breakdown of the energy in the kinetic and potential parts.  The kinetic energy rises as a function of time, as occupations inside the Fermi sphere drop down and momentum distribution develops a tail.  The energy content in that tail depends on resolution, though, that is $\text{d}x$.

In careful examination of Fig.~\ref{fig:fricyesno} and the right panel in Fig.~\ref{fig:av_density}, a slow gradual change in the late-time size and average density can be seen besides oscillations.  These indicate that the system temperature is not completely zero.  Artifacts such as slight temperature and mild oscillations are something that one will have to live with in the collision simulations, with a hope that they might not affect the reaction dynamics much if the reaction energy is elevated and the reaction progresses rapidly.

\section{Next Step: Isospin Dynamics}
 The next natural step in the progression of nuclear NGF investigations is the investigation of collective oscillations in the presence and absence of correlations.  The low-lying and relatively weakly damped nuclear oscillations are those of protons against neutrons.  For study of those oscillations in a systematic way, we need to describe the nuclear symmetry energy in a controlled way.  Contributions to the symmetry energy stem from the Fermi-gas kinetic energy, the correlation energy and Hartree-Fock contribution.  The Hartree-Fock contribution to the net symmetry energy is simply
\begin{equation}
\mathcal{E}_{HF}^a= \int \text{d}x  \; S_{HF}(n) \,  \frac{(n_n-n_p)^2}{n} \, .
\end{equation}
The corresponding contributions to the Hartree-Fock self-energy are then
\begin{equation}
\Sigma_{HF}^{p(n)} = \frac{\partial}{\partial n_{p(n)}}\left [ S_{HF}(n)  \;  \frac{(n_n-n_p)^2}{n} \right]  
= \mp 2 \eta  \, S_{HF} + \eta^2 (n S_{HF}' - S_{HF}) \, ,
\end{equation}
where $\eta = (n_n - n_p)/n$ and the upper r.h.s.\ sign pertains to protons and lower to neutrons.  The correlation contribution stems from asymmetry in the blocking for protons and neutrons and the corresponding difference developing for occupations of neutron and proton states, even when residual interaction is assumed to be isoscalar, cf.~Eq.~\eqref{eq:Vres} and Fig.~\ref{fig:borniso}.  At a later stage this may need to be modified.
 
\section{Conclusions}

We have carried out first steps towards exploring the dynamics of correlated nuclear systems in NGF.  We prepared correlated infinite and finite nuclear systems in 1D through adiabatic switching.  We explored the use of different switching functions, initial conditions and cooling friction, arriving at reasonably stationary and cold states for initiating the dynamics of interest.  In parallel, we developed a suitable combination of mean-field and residual interactions for describing the initial states and the dynamics.

\section{ACKNOWLEDGMENTS}

The authors benefited from discussions with Brent Barker and Hao Lin.  This work was supported by the U.S.\ National Science Foundation under Grant PHY-1520971.

\bibliography{refr}%
\bibliographystyle{aipnum-cp}%

\end{document}